\documentclass[prd,superscriptaddress,amsfonts,amssymb,amsmath,showpacs,twocolumn]{revtex4-2}
\usepackage{bm}
\usepackage{amsfonts}
\usepackage{latexsym}
\usepackage[latin1]{inputenc}
\usepackage{graphicx}
\usepackage{amsmath}
\usepackage{palatino}
\usepackage{mathpazo}
\usepackage{textcomp}
\usepackage{float}
\usepackage{booktabs}
\usepackage{dcolumn}
\usepackage{ragged2e}
\usepackage{hyperref}
\hypersetup{colorlinks,citecolor=blue}
\hypersetup{colorlinks=true,linkcolor=red,filecolor=magenta,    urlcolor=blue}
\usepackage{amsmath}
\usepackage{xcolor}
\usepackage{orcidlink}
\usepackage[caption=false]{subfig}
\usepackage{commath}
\def\jnl@style{\it}
\def\aaref@jnl#1{{\jnl@style#1}}

\def\aaref@jnl#1{{\jnl@style#1}}

\def\aj{\aaref@jnl{AJ}}                   % Astronomical Journal
\def\apj{\aaref@jnl{ApJ}}                 % Astrophysical Journal
\def\apjl{\aaref@jnl{ApJ}}                % Astrophysical Journal, Letters
\def\apjs{\aaref@jnl{ApJS}}               % Astrophysical Journal, Supplement
\def\apss{\aaref@jnl{Ap\&SS}}             % Astrophysics and Space Science
\def\aap{\aaref@jnl{A\&A}}                % Astronomy and Astrophysics
\def\aapr{\aaref@jnl{A\&A~Rev.}}          % Astronomy and Astrophysics Reviews
\def\aaps{\aaref@jnl{A\&AS}}              % Astronomy and Astrophysics, Supplement
\def\mnras{\aaref@jnl{Mon.~Not.~Roy.~Astron.~Soc.}}             % Monthly Notices of the RAS
\def\prd{\aaref@jnl{Phys.~Rev.~D}}        % Physical Review D
\def\plb{\aaref@jnl{Phys.~Lett.~B}}        % Physics Letters B
\def\prc{\aaref@jnl{Phys.~Rev.~C}}  % Physical Review C
\def\prl{\aaref@jnl{Phys.~Rev.~Lett.}}    % Physical Review Letters
\def\qjras{\aaref@jnl{QJRAS}}             % Quarterly Journal of the RAS
\def\skytel{\aaref@jnl{S\&T}}             % Sky and Telescope
\def\ssr{\aaref@jnl{Space~Sci.~Rev.}}     % Space Science Reviews
\def\zap{\aaref@jnl{ZAp}}                 % Zeitschrift fuer Astrophysik
\def\nat{\aaref@jnl{Nature}}              % Nature
\def\aplett{\aaref@jnl{Astrophys.~Lett.}} % Astrophysics Letters
\def\apspr{\aaref@jnl{Astrophys.~Space~Phys.~Res.}} % Astrophysics Space Physics Research
\def\physrep{\aaref@jnl{Phys.~Rep.}}      % Physics Reports
\def\physscr{\aaref@jnl{Phys.~Scr}}       % Physica Scripta
\def\commat{\aaref@jnl{Comm.~Math.~Phys.}}              % Communications in Mathematical Physics
\def\science{\aaref@jnl{Science}}               % Science
\def\cqg{\aaref@jnl{Classical Quant.~Grav.}}            % Classical and Quantum Gravity
\def\jpcs{\aaref@jnl{JPCS}}                                     % Journal of Physics Conference Series
\def\ijmpd{\aaref@jnl{Int.~J.~Mod.~Phys.~D}}                    % International Journal of Modern Physics D
\def\grg{\aaref@jnl{Gen.~Relat.~Gravit.}}               % General Relativity and Gravitation
\def\rpp{\aaref@jnl{Rep.~Prog.~Phys.}}          % Reports on Progress in Physics
\def\npa{\aaref@jnl{Nucl.~Phys.~A}}        % Nuclear Physics A
\def\lrr{\aaref@jnl{Living Rev.~Rel.}}                   % Living reviews in relativity
\def\jcap{\aaref@jnl{J.~Cosmology Astropart.~Phys.}}    % Journal of cosmology and astroparticle physics
\def\rmp{\aaref@jnl{Rev.~Mod.~Phys.}}   %Reviews of modern physics
\def\epjc{\aaref@jnl{Eur.~Phys.~J.~C}}

\allowdisplaybreaks[1]
\begin{document}
%\color{red}
\color{black}       %% For one column
\title{Electrically Charged Quark Stars in $4D$ Einstein-Gauss-Bonnet Gravity}

\author{Juan M. Z. Pretel  \orcidlink{0000-0003-0883-3851}}
 \email{juanzarate@if.ufrj.br}%Lines break automatically or can be forced with \\
 \affiliation{
 Instituto de F\'\i sica, Universidade Federal do Rio de Janeiro,\\
 CEP 21941-972 Rio de Janeiro, RJ, Brazil 
}

\author{Ayan Banerjee \orcidlink{0000-0003-3422-8233}} 
\email{ayanbanerjeemath@gmail.com}
\affiliation{Astrophysics and Cosmology Research Unit, School of Mathematics, Statistics and Computer Science, University of KwaZulu--Natal, Private Bag X54001, Durban 4000, South Africa}

\author{Anirudh Pradhan} \email[]{pradhan.anirudh@gmail.com}
\affiliation{Department of Mathematics, Institute of Applied Sciences and Humanities, GLA University, Mathura-281 406, Uttar Pradesh, India}

%%%%%%%%%%%%%%%%%%%%%%%%%%%%%%%%%%%%%  DATE  %%%%%%%%%%%%%%%%%%%%%%%%%%%%%%%%%%%%

\date{\today}

\begin{abstract}
In this work we study the properties of compact spheres made of a charged perfect fluid with a MIT bag model EoS for quark matter. Considering static spherically symmetric spacetime we derive the hydrostatic equilibrium equations in the recently formulated four dimensional Einstein-Gauss-Bonnet ($4D$ EGB) gravity theory. In this setting, the modified TOV equations are solved numerically with the aim to investigate the impact of electric charge on the stellar structure. A nice feature of $4D$ EGB theory is that the Gauss-Bonnet term has a non-vanishing contribution to the gravitational dynamics in $4D$ spacetime. We therefore analyse the effects of Gauss-Bonnet coupling constant $\alpha$ and the charge fraction $\beta$ on the mass-radius ($M-R$) diagram and also the mass-central density $(M-\rho_c)$ relation of quark stars. Finally, we conclude that depending on the choice of coupling constant one could have larger mass and radius compared with GR and can also be relevant for more massive compact objects due to the effect of the repulsive Coulomb force. 

\end{abstract}

\maketitle

\section{Introduction}

General relativity (GR) is the most successful gravity theory for more than 100 years, almost universally accepted and well confirmed by scrutiny and testing \cite{Will:2014kxa}. Nevertheless, there are a number of unsolved problems of GR both from a theoretical and observational point of view (see Ref. \cite{Coley:2018mzr}) and thereby a number of alternative models have been proposed. Moreover, the current versions of string theory require 10 dimensions total or 11 if you take an extended version known as M-Theory. However, the existence of extra spacetime dimensions beyond the four was initiated by  Kaluza \cite{Kaluza:1921tu} and  Klein \cite{Klein:1926tv}, and now it is known as Kaluza-Klein theory.  In this context, Lovelock (or referred to as Lanczos-Lovelock)  \cite{Lovelock:1971, Lovelock:1972} higher-curvature gravity theory is rather special.  In particular, Lovelock theory is the  most natural extension of GR in higher dimensional spacetimes while keeping the order of the field equations down to second order in derivatives without torsion. A notable fact is that such theory is known to be free of ghosts  \cite{Zwiebach, Zumino} when considering perturbations around flat spacetime.

Among all the classes of Lovelock theory of gravity, the simplest non-trivial Lovelock gravity is the so-called Einstein-Gauss-Bonnet (EGB) gravity, whose Lagrangian is the sum of the curvature scalar with a cosmological constant, while the third term contains the quadratic Gauss-Bonnet (GB) term. The EGB gravity has been widely studied because it is realized in the low-energy limit for strings propagating in curved spacetime \cite{Wiltshire:1985us,Wheeler:1986}. In this theory, the static and spherically symmetric black hole solution was found by Boulware and Deser \cite{Boulware:1985}.
However, in $4D$ spacetime, the GB term is a topological invariant and thus does not contribute to the gravitational dynamics,  except it is coupled to a matter field \cite{Odintsov:2020,Odintsov:2020sqy}.

But, recently an interesting observation has been put forward in reformulating the EGB gravity in $4D$ spacetime such that a non-vanishing effect of the GB Lagrangian can emerge.
The basic idea was to rescale the GB coupling constant $\alpha \to \alpha/(D -4)$ because the presence of an overall factor $(D-4)$ will cancel out, and then taking the limit $D \to 4$ \cite{Glavan:2019inb}. This idea is known to be 4$D$ EGB theory and the GB term produces non-trivial contributions to gravitational dynamics. Such a theory would bypass the conclusions of Lovelock's theorem and avoid the Ostrogradsky instability \cite{Woodard:2015zca}. Research on 4$D$ EGB gravity broadened rapidly among the scientific community, such as black hole solutions \cite{Ghosh:2020syx, EslamPanah:2020hoj,Konoplya:2020juj, Singh:2020xju, HosseiniMansoori:2020yfj, Singh:2020nwo, Wei:2020poh, Yang:2020jno}, a Vaidya-like radiating black hole in Ref. \cite{Ghosh:2020vpc}, black holes coupled to magnetic charge and nonlinear electrodynamics \cite{Abdujabbarov:2020jla, Jafarzade:2020ilt} and charged black hole \cite{Fernandes:2020rpa, Zhang:2020sjh}. Furthermore, deflection of light by black holes \cite{Islam:2020xmy, Jin:2020emq, Kumar:2020sag}, weak cosmic censorship conjecture \cite{Yang:2020czk}, quasi-normal modes \cite{Churilova:2020aca,Mishra:2020gce,Aragon:2020qdc}
and shadow cast by black holes \cite{Konoplya:2020bxa, Guo:2020zmf, Zeng:2020dco,Wei:2020ght} have been fully investigated.  A comprehensive analysis about stellar structure models within this framework have been exhaustively 
studied, see for example \cite{Banerjee:2020stc, Banerjee:2020yhu, Doneva:2020ped, Banerjee:2020dad}. Wormhole and thin-shell wormholes have been studied as well \cite{Jusufi:2020yus, Liu:2020yhu}. 

Associated with the success of this theory there are several criticisms against this model. The most important issue is that despite the fact that the Lagrangian has an overall factor $D-4$, this is not the case for the equation of motion. Therefore, the whole limiting procedure was cast in doubt \citep{Gurses:2020ofy,Ai:2020peo,Shu:2020cjw,Mahapatra:2020rds}. As a result, some proposals have been raised to circumvent the aforementioned shortcomings, but possessing some of the flavor of the original idea. With an additional scalar degree of freedom the regularized theories help us to obtain the field equations in a $4D$ version, for example Kaluza-Klein-reduction procedure of the higher-dimensional EGB theory \cite{Lu:2020iav, Kobayashi:2020wqy}, conformally invariant scalar field equation of motion \cite{Fernandes:2021dsb, Fernandes:2020nbq,Hennigar:2020lsl}, 
and ADM decomposition analysis \cite{Aoki:2020lig}. Thus, it reflects that regularized procedures are not unique. It is however worthwhile to remark that spherically-symmetric $4D$ solutions still remain valid in these regularized theories \cite{Banerjee:2020yhu} as of original prescription presented in \cite{Glavan:2019inb}.

In considering alternative theories of gravity it is necessary to pass constraints obtained from the classical tests of GR at an observational and theoretical level. The best way to check the viability of this theory one may start from strong-field regime \cite{Psaltis:2008bb}. With this point of view compact astrophysical objects, such as neutron stars (NSs), can be considered suitable test-beds in the strong-field regime. NSs are considered to be the most ideal astrophysical laboratories for dense nuclear matter that end their life cycles of massive stars $8 M_{\odot} \lesssim M \lesssim 25 M_{\odot}$ via supernova explosions. The observations made over the electromagnetic spectrum suggest that NSs can have mass between 1-2 $M_{\odot}$ with radius between 10-15 {\rm km} \cite{Ozel:2016oaf,Steiner:2017vmg}. As a result, in a narrow range the central densities are several times higher than nuclear saturation density i.e., $\rho \gtrsim \rho_{\text{nuc}} $ where $\rho_{\text{nuc}} = 2.8 \times 10^{14}$ $\rm g/cm^3$. Such extreme conditions make it impossible to deal with this type of matter in a laboratory conducted  on Earth, and thus we are far away from the comprehensive picture of their internal structure.

On the other hand, measurements of the masses or radii have put a strong constraint on the equation of state (EoS) governing these compact objects, and consequently the interior composition also. For this reason, physicists predict different types of effective models (exotic matter with large strangeness fraction) in order to extract their bulk properties and reveal how matter behaves in their interior. For instance,
``exotic matter'' such as a Bose-Einstein condensate of negative pions ($\pi^{-}$) or negative kaons ($K^{-}$), hot quark-gluon plasma and even cold quark matter  are now so widely  discussed that they are familiar concepts. Among them quark matter in the core of compact objects might be absolutely stable and thus the true ground state of hadronic matter \cite{Itoh,Witten:1984rs}. This possibility was first realized by Witten \cite{Witten:1984rs} and Bodmer \cite{Bodmer:1971we} that compact stars are partially or totally made of quarks. Even more intriguing  the existence of a quark core in a NS is constituted of almost equal numbers of  \textit{up, ~down~ and ~strange} quarks, and a small number of electrons to attain the charge neutrality. Such compact stars have been called strange quark stars (SQS) (shortened as strange stars) and the MIT-bag model is one of the most successful ones for quark confinement. We will discuss below more about MIT-bag model.

So far most of the studies have performed under the assumption of charge neutrality inside a spherical surface.
But, the breakthrough came from several researchers \cite{Olson1,Olson2}. Moreover, the existence of SQS ought to be made of chemically equilibrated strange matter, and it
requires the presence of electrons inside strange stars \cite{Negreiros:2009fd}. In such a system, electrons play 
an important role in producing repulsive force which will add up to the internal pressure allowing more  repulsive force of the system. In 1924, Rosseland \cite{Rosseland} first pointed out the possibility of existing of a self-gravitating star with non-vanishing net charge where the star is modeled by a ball of hot ionized gas. Currently, there are couple of articles showing the existence of charged stars \cite{Lemos:2014lza, Arbanil:2017huq}; however, the charge can be as high as $10^{20}$ Coulomb to bring any significant effect on the mass-radius relation \cite{Ray:2003gt}. This situation has been verified by Ray \textit{et al} \cite{Ray:2006qq} (see Ref. \cite{Panotopoulos:2021cxu} for a more detailed discussion). 

In recent progress, Ivanov \cite{Ivanov:2002jy} assumed a linear EoS to study charged perfect fluid solutions. Motivated by this results, several authors have studied electrically charged fluid spheres for linear and non-linear EoS \cite{Varela:2010mf, Kumar:2018rlo, Nasim:2018ghs, Thirukkanesh:2008xc, Panotopoulos:2019wsy}. Motivated by MIT bag models of strange stars, charged solutions were studied by some authors \cite{Arbanil:2015uoa, Malheiro:2011zz}. Studies on the stability of charged fluid spheres against radial perturbation 
have been done in Ref. \cite{Arbanil:2015uoa}. For in depth discussions about the stability of charged spheres have been found in \cite{Andreasson:2008xw,Sharma:2020ooh,Wang:2020dwm,Goncalves:2020joq}.
In Ref. \cite{Negreiros:2009fd}, authors showed that the electric charge distribution can have a significant impact on the structure of QSs. A large number of analytic solutions to the Einstein-Maxwell system were also investigated in Refs. \cite{Komathiraj:2008em, Takisa:2013tla, Zubair:2020pvg}. 

In light of these fantastic results, in this paper, we will study the effect of electric charge on compact stars assuming the MIT bag model EoS. We consider the internal structure and their physical properties for a specific classes of QSs in the recently proposed $4D$ EGB gravity theory. To do so, in Section \ref{sec2}, we present the field equations for Einstein-Maxwell-Gauss-Bonnet theory and we derive the modified Tolman-Oppenheimer-Volkoff (TOV) equations describing the star interior for a static and spherically symmetric system. In Section \ref{sec3}, we define the appropriate boundary conditions for interior and exterior spacetime in order to solve the stellar structure equations. In Section \ref{sec4} we discuss the EoS concerning MIT-bag quark model. To simplify our calculation we assume that charge density is proportional to the energy density. In this context, we continue our discussion for numerical findings specially focusing on the mass-radius relation and the stability of hydrostatic equilibrium in Section \ref{sec5}. Finally, our conclusions are reported in Section \ref{sec6}. We adopt a geometric unit system, however, we show our results in physical units for comparison purposes.

%%%%%%%%%%%%%%%%%%%%%%%%%%%%%%%%%%%%%%%%%%%%%%%%%%%%%%
 \section{Basic construction of charged stellar model 
 in  4$D$ EGB gravity} \label{sec2}
%%%%%%%%%%%%%%%%%%%%%%%%%%%%%%%%%%%%%%%%%%%%%%%%%%%%%
 
We start from the action of the Einstein-Maxwell-Gauss-Bonnet theory in $4D$ spacetime introduced in Ref. \cite{Glavan:2019inb} after the rescaling of Gauss-Bonnet constant $\alpha$. Expressed in an explicit manner, the action in $D$ spacetime takes the following form,
\begin{eqnarray}\label{action}
	\mathcal{I}_{A} &=& \frac{1}{16 \pi G}\int d^{D}x\sqrt{-g}\left[ R +\frac{\alpha}{D-4} \mathcal{G} \right] \nonumber \\
	&&+ \int d^{D}x\sqrt{-g} \mathcal{L}_m.
\end{eqnarray}
where $\alpha$ is a coupling constant with dimensions of length squared and the Gauss-Bonnet invariant is defined as
\begin{equation}
\mathcal{G} = R^{\mu\nu\rho\sigma} R_{\mu\nu\rho\sigma}- 4 R^{\mu\nu}R_{\mu\nu}+ R^2\label{GB},
\end{equation}
where $R_{\mu \nu \rho \sigma}$ is the Riemann curvature tensor, $R_{\mu \nu}$ is the Ricci curvature tensor and $R^{2}$ is the squared of the scalar curvature.

The field equations in $D$ dimensional spacetime can be obtained from variation of the action (\ref{action}) with respect to metric tensor $g_{\mu \nu}$, namely
\begin{equation}\label{GBeq}
R_{\mu\nu}-\frac{1}{2}R~ g_{\mu\nu}+\frac{\alpha}{D-4} H_{\mu\nu}=  8 \pi  T_{\mu\nu},
\end{equation}
where $H_{\mu\nu}$  is the Lanczos tensor with the following expressions
\begin{eqnarray}
H_{\mu\nu} &=& 2\Bigr( R R_{\mu\nu}-2R_{\mu\sigma} {R}{^\sigma}_{\nu} -2 R_{\mu\sigma\nu\rho}{R}^{\sigma\rho} - R_{\mu\sigma\rho\delta}{R}^{\sigma\rho\delta}{_\nu}\Bigl) \nonumber \\
&& - \frac{1}{2}~g_{\mu\nu}~\mathcal{L}_{\text{GB}},\label{FieldEq}
\end{eqnarray}
and $T_{ \mu \nu}$ is the energy-momentum tensor of the matter field and which can be calculated by
\begin{equation}
T_{ \mu \nu} =-\frac{2}{\sqrt{-g}}\frac{\delta(\sqrt{-g}\mathcal{L}_{ m})}{\delta g^{ \mu \nu}}.
\end{equation}
The third term on the left-hand side of Eq. (\ref{GBeq}) is the quadratic Gauss-Bonnet term with the property that it contains
at most second-order derivatives. As a result, the above theory has a non-trivial contribution to the gravitational dynamics by taking the limit $D \to 4$.

In the present study we assume that the energy-momentum tensor $T_{\mu\nu}$ is a sum of two terms, i.e. $T_{\mu\nu}= E_{\mu\nu} + M_{\mu\nu}$ in Eq. (\ref{GBeq}). The first part $E_{\mu\nu}$ associates with electromagnetic energy-momentum tensor,
\begin{equation}\label{ElectromagEMT}
E_{\mu\nu}=\frac{1}{4\pi}\left(F_{\mu}\hspace{0.1mm}^{\gamma}F_{\nu\gamma}-
\frac{1}{4}g_{\mu\nu} F_{\gamma\beta}F^{\gamma\beta}\right).
\end{equation} 
where $F_{\mu\nu}$ is the electromagnetic tensor. The
electromagnetic field satisfies the Maxwell equations:
\begin{equation}\label{mfe}
\left[ \sqrt{-g} F^{\mu\nu}  \right]_{,\nu} = 4\pi j^\mu \sqrt{-g}.
\end{equation}
The electromagnetic $D$-current is given by $J^{\mu} = \rho_{ch} u^{\mu}$, where $\rho_{ch}$ is the electric charge density. Moreover, the tensor term $M_{\mu\nu}$ corresponds to an isotropic perfect fluid, whose energy-momentum tensor has the following form 
\begin{equation}\label{pf}
M_{\mu\nu}=(\rho+ P)u_{\mu}u_{\nu}+ Pg_{\mu\nu},
\end{equation}
where $P$ is the pressure, $\rho$ the energy density of matter, and $u_{\nu}$ is the $D$-velocity of the fluid satisfying the normalization condition $u_{\nu}u^{\nu} = -1$.

For our purpose, we consider a static and spherically symmetric metric describing the interior spacetime of a compact star in $D$ dimensions, namely 
\begin{equation}\label{metric}
ds^2_D = - e^{2\Phi(r)}dt^2 + e^{2\lambda(r)} dr^2 + r^2d\Omega^2_{D-2}, 
\end{equation}
\noindent
where $d\Omega^2_{D-2}$ represents the metric on the surface of the $(D-2)$-sphere, given by
\begin{align}\label{D2sphere}
    d\Omega^2_{D-2} =&\ d\theta_1^2 + \sin^2\theta_1d\theta_2^2 + \sin^2\theta_1\sin^2\theta_2 d\theta_3^2  \nonumber  \\
    &+ \cdots + \left( \prod_{j=1}^{D-3} \sin^2\theta_j \right)d\theta_{D-2}^2,
\end{align}
with the metric functions $\Phi$ and $\lambda$ depending on $r$ alone. For this line element it also follows that the only non-vanishing component of the Maxwell strength tensor is $F^{01} = -F^{10}$, and which is a function of $r$, only. The other components of electromagnetic tensor are identically zero. Consequently, Maxwell's Eq. (\ref{mfe}) leads to the following expression
\begin{equation}
E(r) = F^{01}(r) =\frac{4\pi}{r^2} e^{-(\Phi+\lambda)}  \int^r_0r^{D-2} e^{\lambda} {\rho_{ch}} ~dr ,
\end{equation}
and hence the electric charge within a sphere of radius $r$ can be written in a suggestive form by defining the charge function $q(r)$ as
\begin{equation}\label{Q1}
  \frac{d q(r)}{dr}  = 4 \pi r^{D-2} \rho_{ch} e^{\lambda} \, .
\end{equation}

In the limit $D \to 4$, combining the line element (\ref{metric}) together with the energy-momentum tensors (\ref{ElectromagEMT}) and (\ref{pf}), the $00$ and $11$ components of the field equations (\ref{GBeq}) can be explicitly written as
\begin{eqnarray}
&&\frac{\alpha(1-e^{-2\lambda})}{r^3}\left[4\lambda ' e^{-2\lambda}-\frac{(1-e^{-2\lambda})}{r}\right]
\nonumber \\
&&\quad +e^{-2\lambda}\left({2\lambda' \over r}-{1\over r^2}\right) + {1\over r^2} =  8 \pi \Big( \rho +{q^2 \over 8\pi r^{4}}\Big),  \hspace{0.6cm}  \label{DRE1}  \\
&& \frac{\alpha(1-e^{-2\lambda})}{r^3}\left[4\Phi ' e^{-2\lambda}+\frac{(1-e^{-2\lambda})}{r}\right] \nonumber \\
&& \quad +e^{-2\lambda}\left({2\Phi' \over r}+{1\over r^2}\right)-{1\over r^2}= 8 \pi\Big( P- {q^2 \over 8\pi r^{4}}\Big).  \hspace{0.6cm}  \label{DRE2} 
\end{eqnarray}
In addition, the covariant conservation of the energy-momentum tensor, which can be derived from the contracted Bianchi identities (i.e., $\nabla_{\mu}T^{\mu\nu} $ = 0), provides
\begin{eqnarray}
 \frac{dP}{dr} = - (\rho + P) \frac{d\Phi}{dr}+\frac{q q'}{4 \pi r^4}.  \label{DRE3}
\end{eqnarray} 

In order to solve this set of differential equations, one begins with a more familiar form of mass function $m(r)$ through the following relation 
\begin{eqnarray}\label{int}
e^{-2\lambda} = 1+\frac{r^2}{2\alpha} \left[1-\sqrt{1+4\alpha \left(\frac{2m}{r^3} - \frac{q^2}{r^4}\right)}\right] .
\end{eqnarray}
Note that in the limit $\alpha \to 0$ or large $r$, the solution (\ref{int}) behaves asymptotically as
\begin{equation}
\lim_{\alpha\rightarrow 0} e^{-2\lambda} = 1-\frac{2m}{r}+\frac{q^2}{r^2}+\frac{(q^2-2mr)^2}{r^6}\alpha + \cdots .
\end{equation}
We can now rewrite Eq. (\ref{DRE1}) in terms of the mass parameter $m(r)$, this is,
\begin{equation}\label{dmel}
  \frac{dm}{dr} = 4\pi r^2 \rho
  +\frac{q}{ r}\frac{dq}{dr} , 
\end{equation}
which clearly exhibits the same as the corresponding equations in GR, and without influence of coupling constant $\alpha$, as measured in the star's frame.

Finally, the substitution of Eq. (\ref{Q1}) and the conservation equation (\ref{DRE3}) into Eq. (\ref{DRE2}), it yields
\begin{eqnarray}\label{tov}
{dP \over dr} &=& \frac{(P +\rho) \left[r^3 \left(\Lambda + 8\pi \alpha  P - 1\right)- 2\alpha m \right]}{r^2 \Lambda \left[r^2 \left(\Lambda - 1\right)- 2\alpha \right]} \nonumber \\
&&+\frac{q}{4 \pi  r^4} \frac{dq}{dr}, \label{e2.11}
\end{eqnarray}
where
\begin{eqnarray}
    \Lambda^2 \equiv 1+4\alpha \left(\frac{2 m}{r^3}-\frac{q^2}{r^4}\right).
\end{eqnarray}
Eq. (\ref{tov}) is the modified TOV equation, describing the hydrostatic equilibrium of relativistic stars in $4D$ EGB theory. Note that when $\alpha \rightarrow 0$ the above  equation reduce to the TOV equation for electrically charged fluid spheres in GR.

We arrive at six unknown functions, namely, $\Phi(r)$,  $m(r)$, $q(r)$, $\rho(r)$, $P(r)$ and $\rho_{ch}(r)$, with four differential equations: (\ref{Q1}), (\ref{DRE3}), (\ref{dmel}) and (\ref{tov}). Therefore, we need a suitable assumption that reduces the number of unknown functions, see for instance some prescriptions in Refs. \cite{Arbanil:2015uoa, Arbanil:2013pua} for charged fluid spheres in GR. In the following discussion we are free to specify two of the six unknowns; in this treatment we assume an EoS relating the pressure with the energy density of the fluid, and then a relation between the charge distribution and the mass density. Using this assumption we are able to solve the structure equations numerically with some appropriate boundary conditions. We should remark that the metric potential $\lambda(r)$ is then determined by means of Eq. (\ref{int}).

%%%%%%%%%%%%%%%%%%%%%%%%%%%%%%%%%%%%%%%%%%%%%%%%%%%%%%%%%%%%%%

\section{Boundary conditions and the exterior vacuum region to the star} \label{sec3}

To get a stellar structure, we perform a numerical integration from the center at $r= 0$ toward the surface of the star where the pressure vanishes (i.e., when $P(r =R) = 0$). In that regard, we need to establish appropriate boundary conditions for the sought solutions. As already mentioned, given an EoS and a charge density profile, the unknown variables to be determined are $m$, $q$ and $\rho$ through Eqs. (\ref{Q1}), (\ref{dmel}) and (\ref{tov}). Accordingly, we must set the following boundary conditions to maintain regularity at the origin,
\begin{align}\label{BC}
 m(r=0) &= 0,  &  q(r=0) &= 0,  &  \rho(r=0) &= \rho_c, 
\end{align}
where $\rho_c$ is the central energy density. Thus, a set of specific values of $\rho_c$ allows us to obtain a family of compact stars in $4D$ EGB gravity.

Furthermore, to compute the metric function $\Phi$, one needs an additional boundary condition in order to solve Eq.  (\ref{DRE3}). This means that, at the surface of the charged star, the inner and outer potential functions are related through the equality:
\begin{align}\label{intSur}
  e^{2\Phi(R)} &= e^{-2\lambda(R)}  \nonumber  \\
  &= 1+\frac{R^2}{2\alpha} \left[1-\sqrt{1+4\alpha \left(\frac{2 M}{R^3}-\frac{Q^2}{R^4}\right)}\right] ,
\end{align}
with $M = m(r=R)$ and $Q = q(r=R)$ being the total mass and the total charge of the star, respectively. The outer solution corresponds to the Reissner-Nordstr\"{o}m spacetime, if we take the limit $\alpha \to 0$ \cite{Fernandes:2020rpa}.

%%%%%%%%%%%%%%%%%%%%%%%%%%%%%%%%%%%%%%%%%%%%%%%%%%%%%%%%%%%%%%
\section{Equation of state for quark matter and the charge density profile} \label{sec4}
%%%%%%%%%%%%%%%%%%%%%%%%%%%%%%%%%%%%%%%%%%%%%%%%%%%%%%%%%%%%

\subsection{Equation of state}

In this section, we discuss the quark matter model that we are going to use in our analysis. The EoS of quark matter plays a crucial role for determining the star's structure at supranuclear density (above the nuclear density: $\rho_{\rm nuc} = 2.8 \times 10^{14}$ $\rm g/cm^3$) and high temperature ($T \sim 10 ~{\rm MeV}$). Its great success at very high energies has triggered many theoretical investigations both on the modeling of the EoS of quark matter which incorporate basic features of Quantum Chromodynamics (QCD) and on the phenomenological implications related to measurements of masses and radii of compact stars. However, we are still far away from the exact EoS of quark matter.

As a consequence, the MIT bag model is the simplest phenomenological model for quark matter. Such model has been developed to describe hadron properties in terms of quarks. The range in which quarks are confined is called a bag and the energy per unit volume to form it is called a bag pressure, $ {\rm B_{bag}}$. Moreover, the quarks are free inside the bag and are forbidden to reach out. In its simplest form, the EoS of such matter is obtained from the relation
\begin{eqnarray}\label{Prad1}
 P = \dfrac{1}{3}\left(\rho -4B\right),
\end{eqnarray}
where $P$ and $\rho$ represent the pressure
and the energy density of the fluid, respectively, and the parameter $B$ is the bag constant. Note that the external pressure acting on a bag filled with quarks vanishes for $\rho = 4B$. It has been found that the accepted values of Bag constant $B$ lies within the range of  $57 \leq B \leq 92$ MeV/fm$^3$ \cite{Burgio:2018mcr, Blaschke:2018mqw}. In the present work, we will use $B = 57\ \rm MeV/fm^3$. In GR, this value is commonly used because it produces maximum-mass configurations of about $2M_\odot$. Since the aim is to study the degree of modification with respect to Einstein's theory, in this work we will assume the same value for $B$.

\subsection{The charge density relation}

Since our purpose is to investigate the effects of the electric charge on stellar structure, we need to specify the charge density as well. Followed by the discussion in Refs. \cite{Ray:2003gt, Panotopoulos:2019wsy, Arbanil:2013pua}, we assume that the charge density is proportional to the energy density, i.e.
\begin{equation}
  \rho_{ch} = \beta \rho ,
\end{equation}
where $\beta$ is a charge parameter and it measures the amount of charge within the fluid sphere. This assumption is reasonable, as more material is expected to bring along a higher amount of electric charge. As in Ref. \cite{Panotopoulos:2019wsy}, in our work we will use moderate values for $\beta$ which generate appreciable changes in the mass-radius diagrams. Furthermore, Arba{\~n}il and collaborators \cite{Arbanil_2014, Arbanil:2013pua} have shown that the largest value for $\beta$ is 0.99 in order to avoid numerical convergence troubles. Finally, it is worth mentioning that there are other models in the literature to describe electrically charged quark stars, see for instance Refs. \cite{Arbanil:2015uoa, Felice1999} where it is assumed that the charge is proportional to spatial volume.

%%%%%%%%%%%%%%%%%%%%%%%%%%%%%%FIGURES%%%%%%%%%%%%%%%%%%%%%%%%%%%%

\begin{figure*}
 \includegraphics[width=8cm]{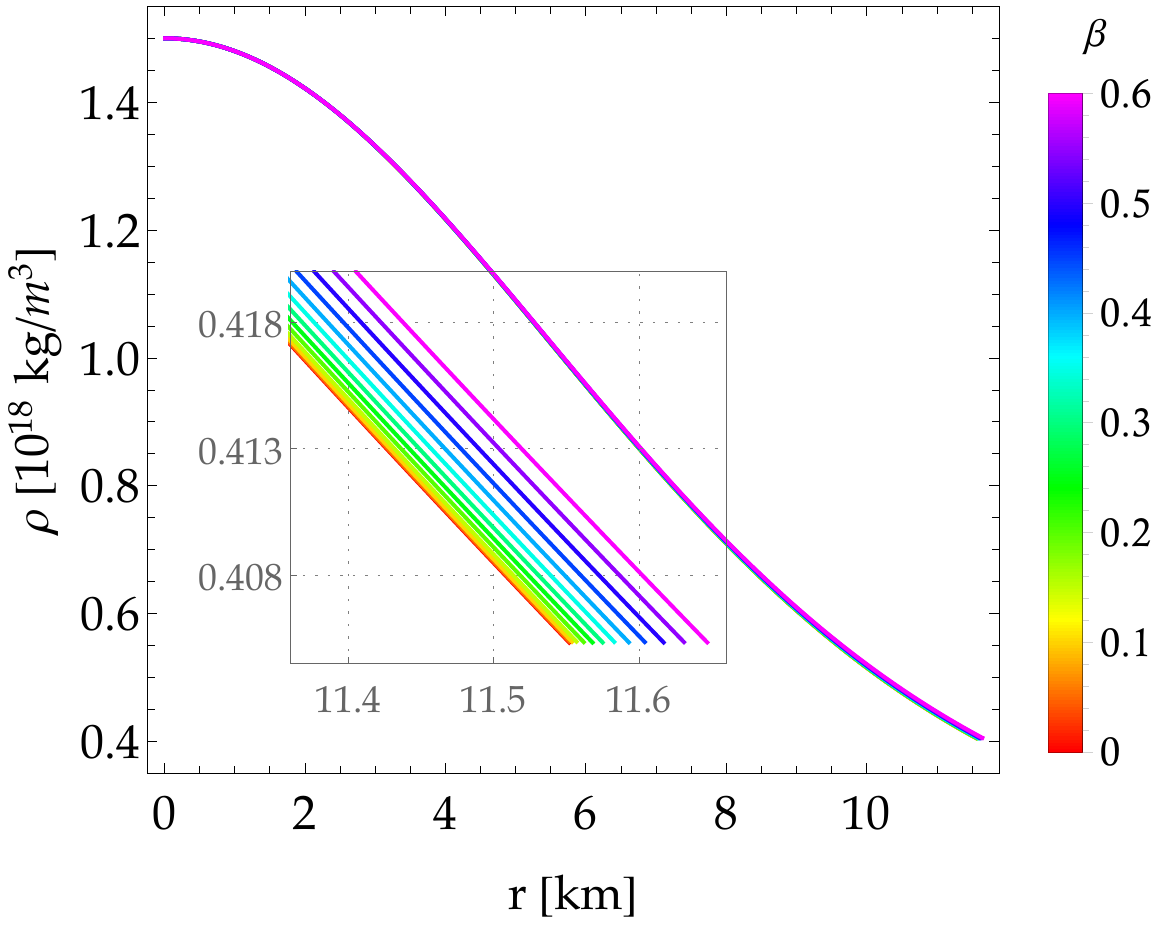}\
 \includegraphics[width=8cm]{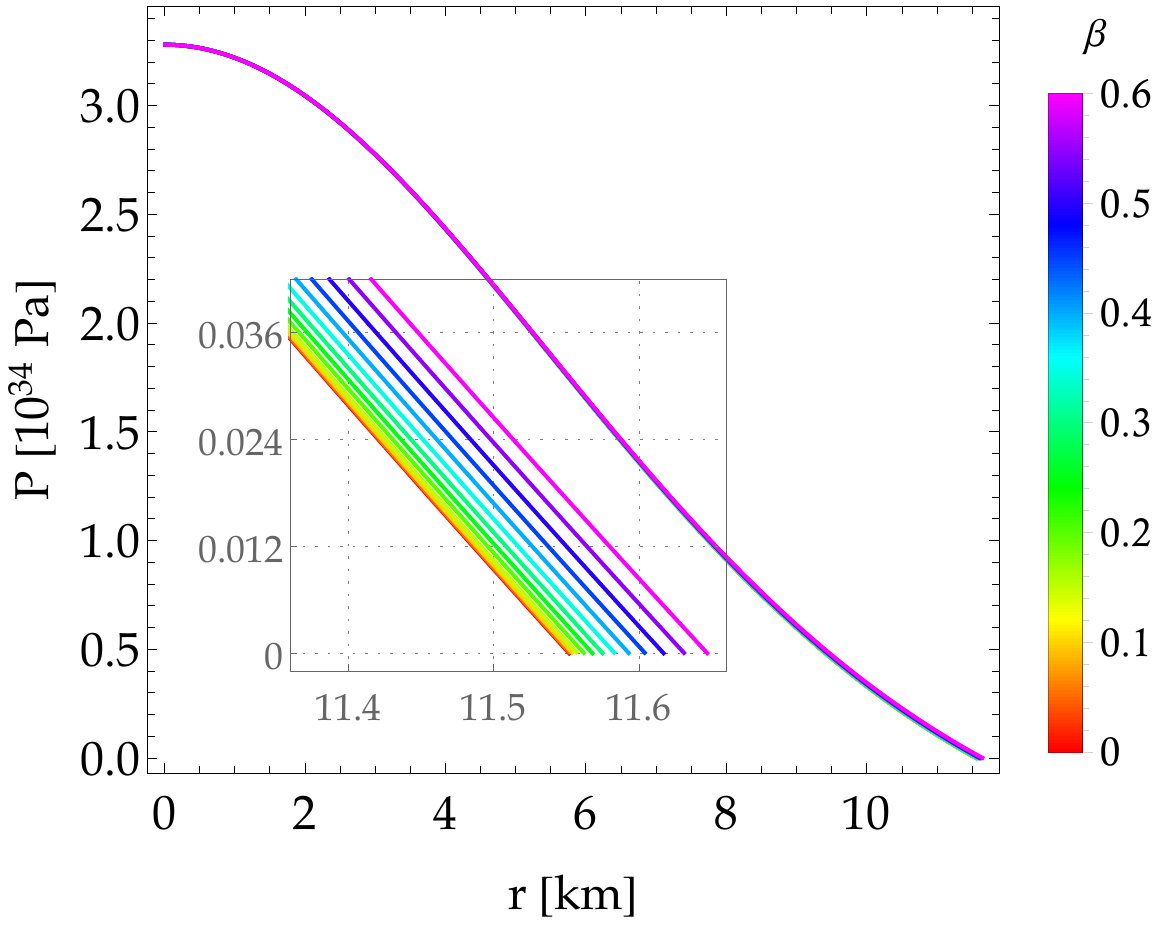}
 \caption{\label{figure1} Numerical solution of the system of differential equations (\ref{Q1}), (\ref{dmel}) and (\ref{tov}) with boundary conditions (\ref{BC}). Radial behavior of the mass density (left panel) and pressure (right panel) for a central density $\rho_c = 1.5 \times 10^{18}\ \rm kg/m^3$ with EoS (\ref{Prad1}) within the framework of $4D$ EGB gravity for $\alpha = 4 \rm km^2$. The color scale on the right side of each plot indicates different values for the charge parameter $\beta$. }  
\end{figure*}

\begin{figure*}
 \includegraphics[width=8cm]{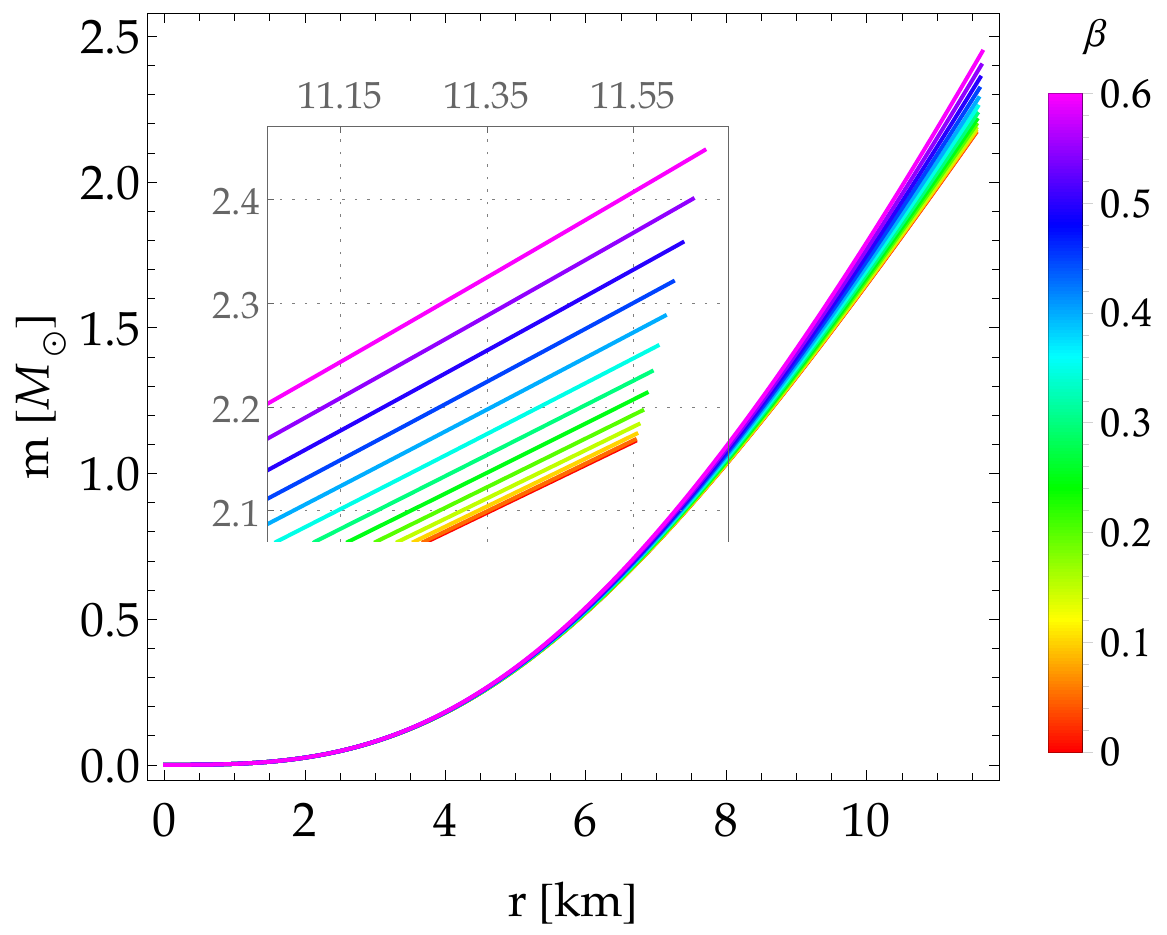}\
 \includegraphics[width=8cm]{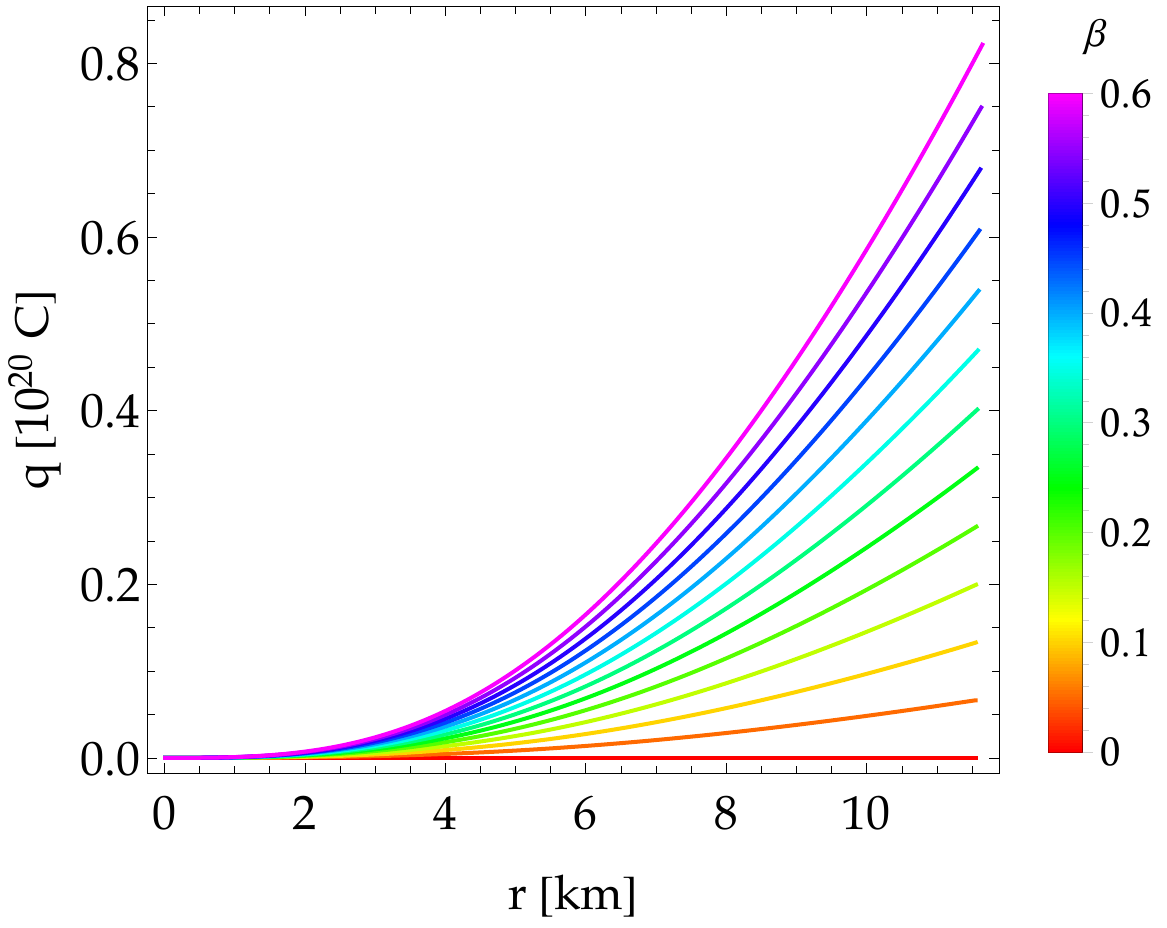}
 \caption{\label{figure2} Mass parameter (left panel) and electric charge (right panel) as functions of the radial coordinate. The results are for the same solution as in Fig. \ref{figure1}. It can be observed that the electric charge increases both the radius and the mass (on the surface) as $\beta$ becomes more positive. }  
\end{figure*}

\begin{figure*}
 \includegraphics[width=8.4cm]{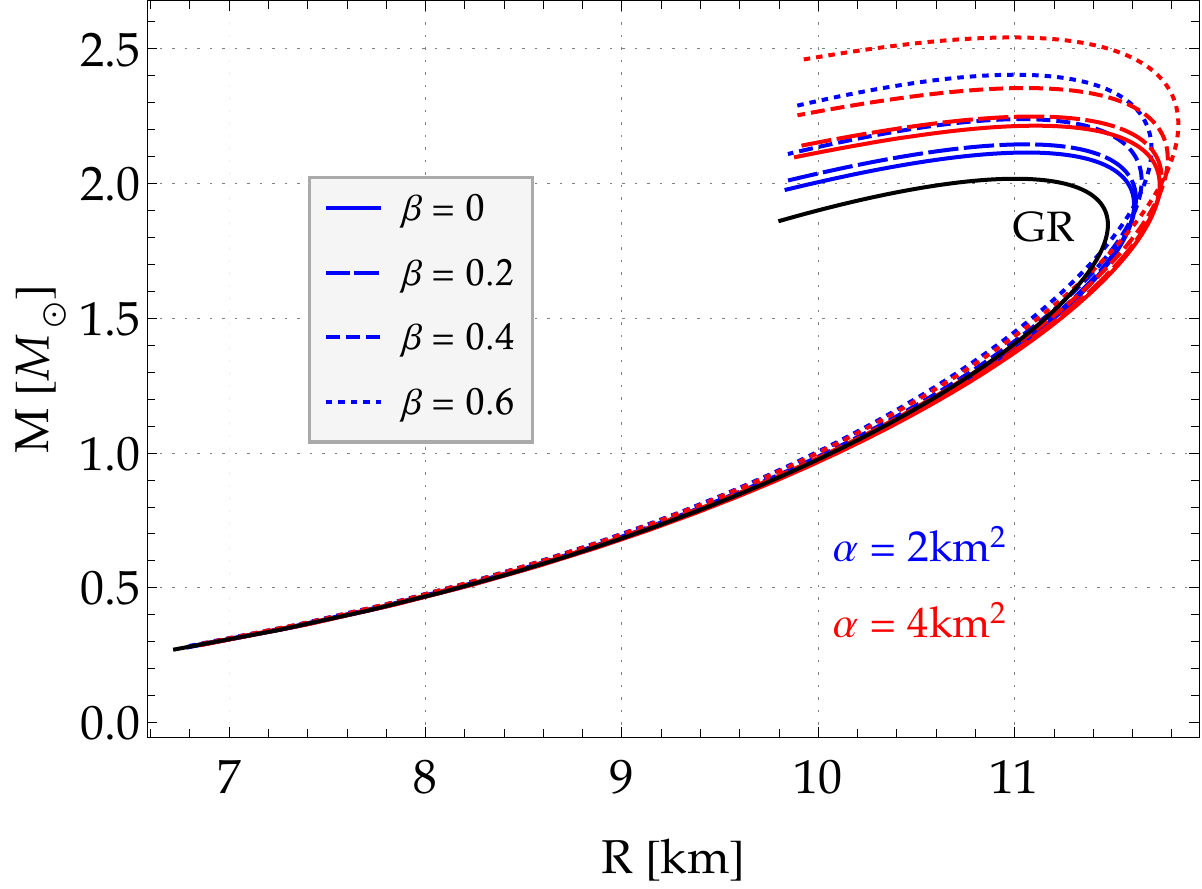}\
 \includegraphics[width=8.41cm]{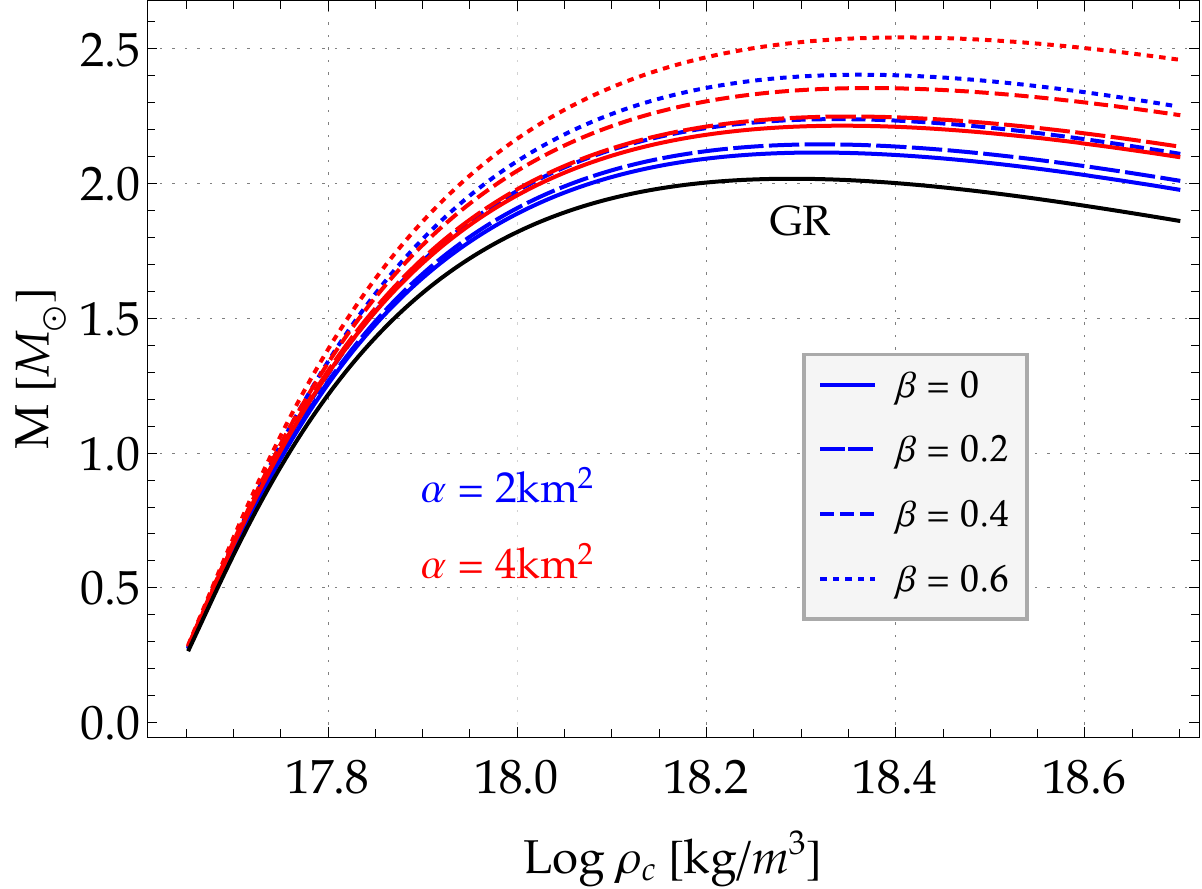}
 \caption{\label{figure3} Mass-radius diagram (left panel) and mass-central density relation (right panel) for charged quark stars in $4D$ EGB gravity. The blue and red lines correspond to $\alpha = 2 \rm km^2$ and $\alpha = 4 \rm km^2$, respectively. Moreover, different styles of the curves stand for different values of the charge parameter $\beta$, and the non-charged GR case is shown in both plots as a benchmark by the black solid line. We note that the gravitational mass of quark stars undergoes slight changes at low central densities and the most significant deviations occur at higher densities due to the charge and GB coupling constant. The stable quark stars on the mass-central density curves in the right panel are found in the region where $dM/d\rho_c >0$.}   
\end{figure*}

\begin{figure*}
 \includegraphics[width=8.4cm]{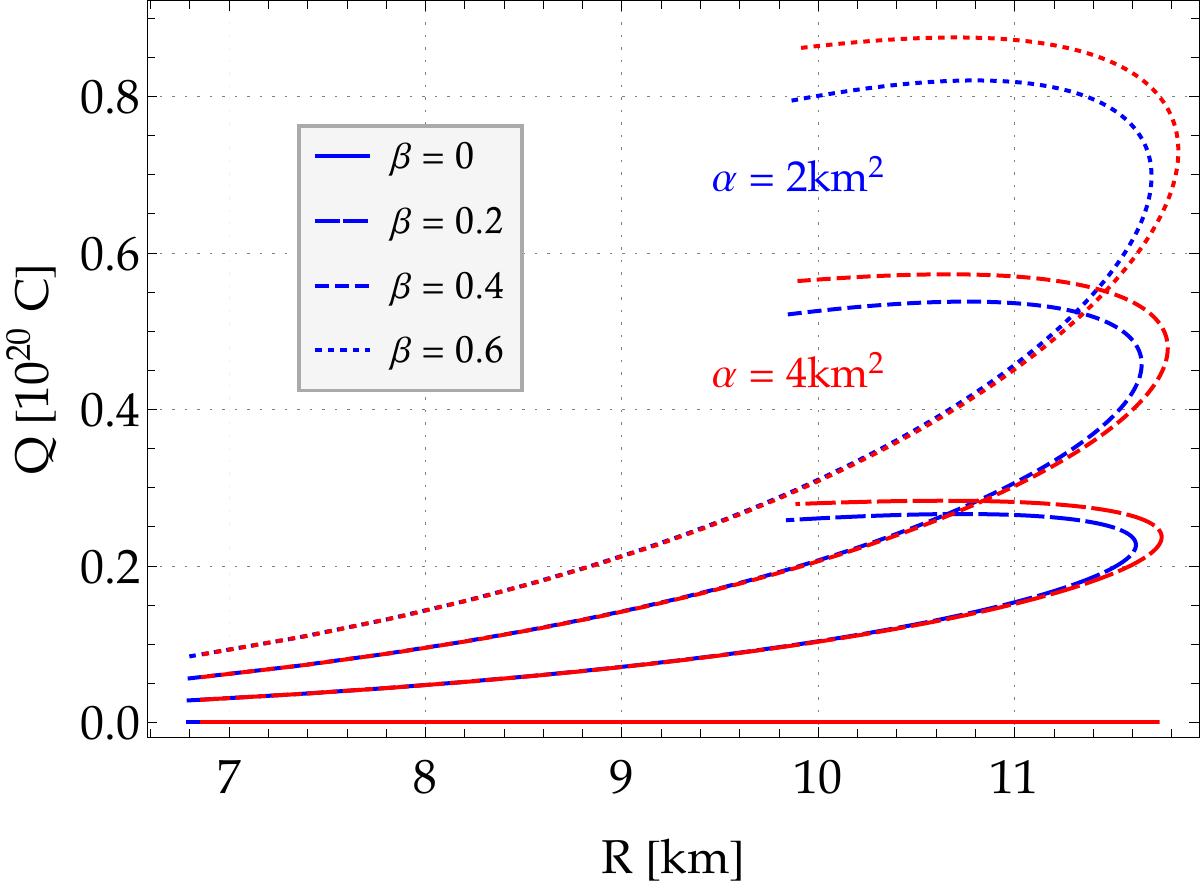}\
 \includegraphics[width=8.41cm]{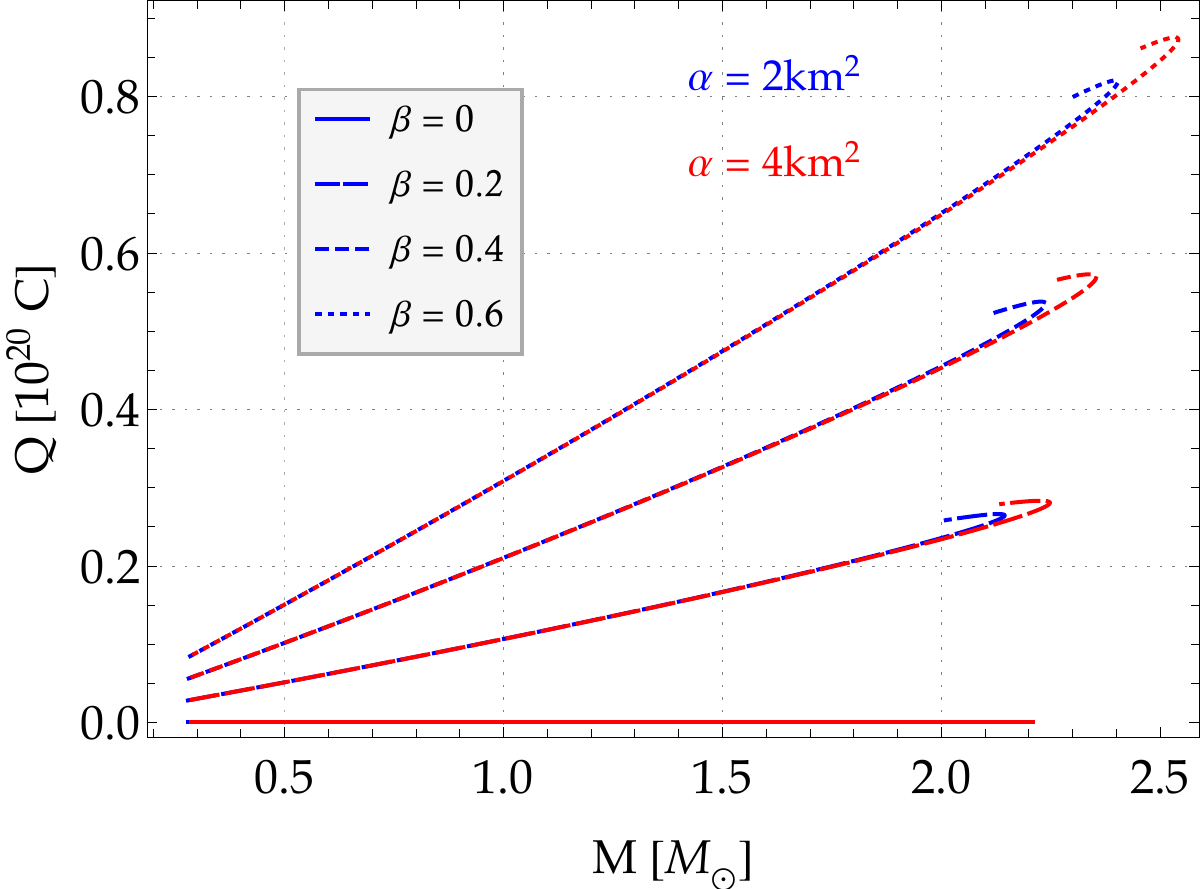}
 \caption{\label{figure4} The total charge as a function of the radius (left panel) and of the mass (right panel). The results for the several values of the parameter $\beta$ as in Fig. \ref{figure3} are shown. It is observed that the electric charge undergoes relevant changes due to the GB term only in the high-mass region. }  
\end{figure*}

\begin{figure}
 \includegraphics[width=8.0cm]{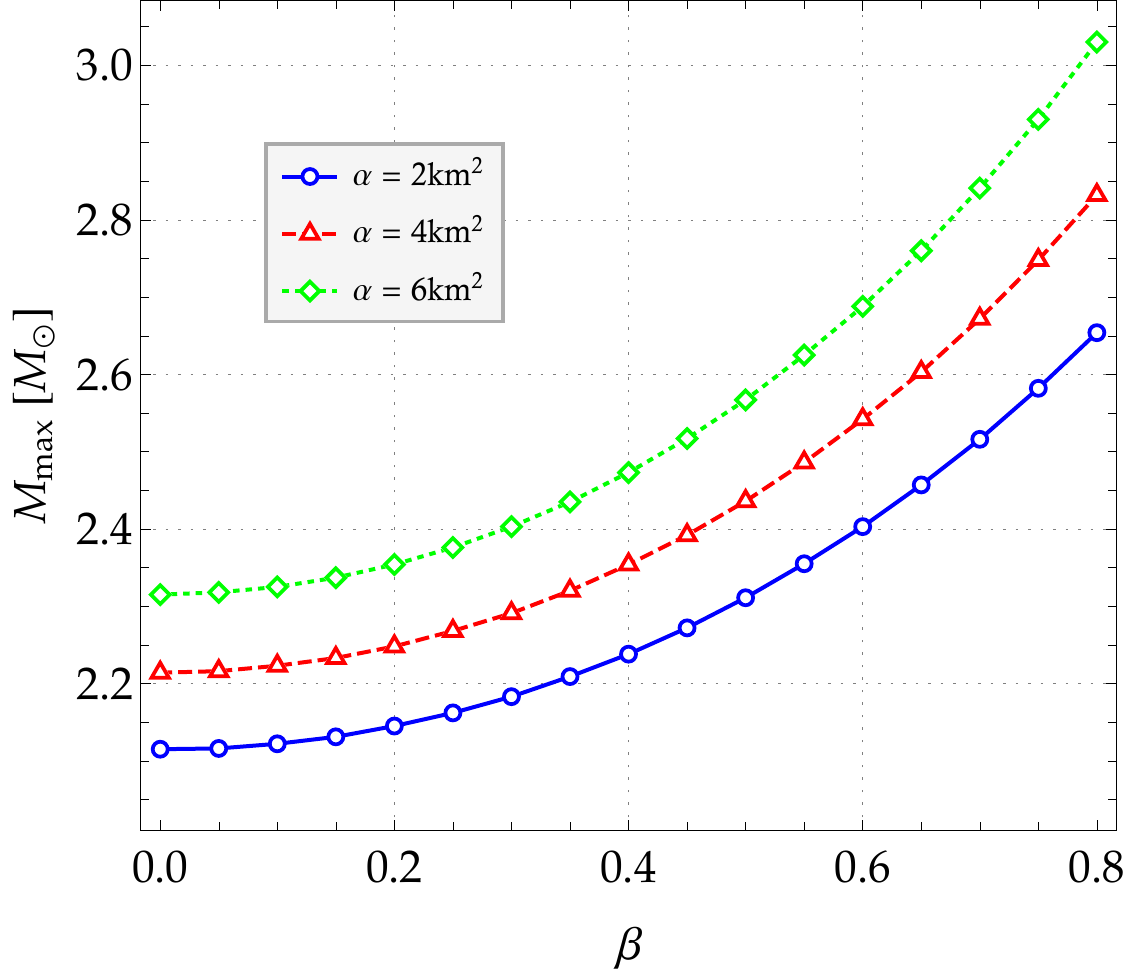}
 \caption{\label{figure5} Maximum mass as a function of the charge parameter $\beta$ for three values of the GB coupling constant, where a quadratic behavior can be observed. }  
\end{figure}

\section{Numerical results and discussion} \label{sec5}

Given a specific value of the coupling constant $\alpha$ and charge parameter $\beta$, the system of modified TOV equations (\ref{Q1}), (\ref{dmel}) and (\ref{tov}) with boundary conditions (\ref{BC}) is numerically integrated from the origin at $r= 0$ up to the stellar surface where the pressure vanishes at $r= R$. In particular, for a central density $\rho_c = 1.5 \times 10^{18}\ \rm{kg}/\rm{m}^3$ with EoS (\ref{Prad1}), Fig. \ref{figure1} displays the mass density and pressure as functions of the radial coordinate for $\alpha= 4\rm km^2$ and different values of $\beta$. Furthermore, according to Fig. \ref{figure2}, the interior structure of a quark star in $4D$ EGB gravity is modified due to the charge parameter $\beta$ and, therefore, both radius and total mass increase as $\beta$ becomes more positive. Note that the mass function and electric charge are increasing quantities as we approach the surface, as expected.

The total gravitational mass and total electric charge of a particular star are determined at the surface, that is, $M = m(R)$ and $Q = q(R)$, respectively. For the central density considered in the previous paragraph, Table \ref{table1} shows some numerical values for the global properties of quark stars for a fixed value of $\alpha$, and increasing $\beta$. Meanwhile, in Table \ref{table2} we have fixed the value of $\beta$ and varied the GB coupling constant. These results indicate that the basic properties of a compact star such as radius and mass undergo considerable changes due to the inclusion of electric charge as well as the Gauss-Bonnet term.

The mass-radius diagrams and mass-central density relations for charged quark stars in $4D$ EGB gravity are presented in Fig. \ref{figure3}, where two values of $\alpha$ have been considered. The results corresponding to Einstein gravity have been included by a black solid line for comparison reasons. Such plots reveal that both mass and radius vary slightly from GR in the low-mass region. Nonetheless, the mass-radius curves exhibit significant changes with respect to the general relativistic counterpart for higher central densities (i.e., close to the maximum-mass point). In other words, the maximum mass of quark stars can be increased by means of the GB constant (with positive values) as well as by the effect of electric charge. This could help explain the existence of supermassive compact stars observed in nature and that Einstein's theory fails to predict even using exotic equations of state.

A traditional technique widely used in the literature to indicate the onset of instability is the $M(\rho_c)$ method, namely, a turning point from stability to instability occurs when $dM/d\rho_c =0$. Therefore, according to this criterion, the stable quark stars on the mass versus central density curves in the right panel of Fig. \ref{figure3} are found in the region where $dM/d\rho_c >0$. Here we must point out that this condition is necessary but not sufficient for stability analysis. A more suitable approach would be to calculate the frequency of the oscillation modes when the star is subjected to radial perturbations. We will leave this study for future work.

Furthermore, we can investigate how the total charge behaves in terms of radius and mass. According to Fig. \ref{figure4}, the charge increases with increasing $\beta$, as expected. However, the changes in electrical charge associated with the Gauss-Bonnet term are more relevant for high values of mass and radius. Finally, we briefly analyze how the maximum mass changes as we vary the charge parameter $\beta$ in the mass-radius curves shown in Fig. \ref{figure3}. As illustrated in Fig. \ref{figure5} for three values of the coupling constant $\alpha$, the maximum mass has a quadratic behavior with $\beta$. Notice that in this plot the main effect of both parameters $\alpha$ and $\beta$ on the maximum mass of charged quark stars can be better observed.

\begin{table}
\caption{\label{table1} 
Charged quark stars with central mass density $\rho_c = 1.5 \times 10^{18}\ \rm{kg}/\rm{m}^3$ and EoS (\ref{Prad1}) in $4D$ EGB gravity for $\alpha = 4 \rm km^2$ and several values of the charge parameter $\beta$. The radial behavior of the mass density, pressure, mass parameter and electric charge of these stars is shown in Figs. \ref{figure1} and \ref{figure2}. }
\begin{ruledtabular}
\begin{tabular}{cccc}
$\beta$  &  $R$  [\rm{km}]  &  $M$ [$M_\odot$]  &  $Q$ [$10^{19}$ C]  \\
\colrule
0  &  11.552  &  2.167  &  0  \\
0.1  &  11.555  &  2.174  &  1.328  \\
0.2  &  11.562  &  2.197  &  2.662  \\
0.3  &  11.575  &  2.234  &  4.010  \\
0.4  &  11.593  &  2.287  &  5.379  \\
0.5  &  11.617  &  2.358  &  6.777  \\
\end{tabular}
\end{ruledtabular}
\end{table}

\begin{table}
\caption{\label{table2} 
Properties of charged quark stars with central mass density $\rho_c = 1.5 \times 10^{18}\ \rm{kg}/\rm{m}^3$ and EoS (\ref{Prad1}) in $4D$ EGB gravity for $\beta = 0.2$ and different values of the GB coupling constant. }
\begin{ruledtabular}
\begin{tabular}{cccc}
$\alpha$ [$\rm km^2$]  &  $R$  [\rm{km}]  &  $M$ [$M_\odot$]  &  $Q$ [$10^{19}$ C]  \\
\colrule
1.0  &  11.360  &  2.063  &  2.478  \\
2.0  &  11.428  &  2.108  &  2.539  \\
3.0  &  11.496  &  2.152  &  2.601  \\
4.0  &  11.562  &  2.197  &  2.662  \\
5.0  &  11.628  &  2.241  &  2.724  \\
\end{tabular}
\end{ruledtabular}
\end{table}\

\section{Conclusions} \label{sec6}

Among the higher curvature gravitational theories, the recently formulated $4D$ EGB gravity has been widely studied because the GB term can yield a non-trivial contribution to the gravitational dynamics even in $4D$ spacetime. In such a context, there are several solutions ranging from cosmological to astrophysical applications. Nevertheless, there are still many aspects of this theory that require further study. Motivated by the importance of compact stars, here we have investigated the impact of electric charge on quark stars with MIT bag model EoS within the framework of $4D$ EGB gravity. In other words, we obtained solutions for spherically symmetric distributions of charged perfect fluid, where the distribution of electric charge inside the sphere is proportional to the energy density.

We have numerically solved the modified TOV equations for charged spheres that describe the dense matter inside QSs and obtained the physical quantities for some values of $\alpha$ and $\beta$. Here, we restricted our study for positive values of  $\alpha$, since only in this case more massive stellar configurations could be possible than the general relativistic counterpart, see Ref. \cite{Banerjee:2020yhu}. For a given central density, we have found that the basic properties of a compact star such as mass and radius undergo relevant changes due to the inclusion of electric charge as well as the GB term.

Furthermore, our results reveal that the $M -\rho_c$ curves are almost indistinguishable from GR in the low-mass region, whereas for higher central densities such curves exhibit significant changes with respect to the GR counterpart. We have determined the maximum-mass values of QSs and, as a consequence, we obtained masses larger than 2$M_{\odot}$. From this perspective, one could expect the existence of supermassive compact stars in EGB gravity compared to corresponding stars in Einstein gravity. Finally, we have investigated the stellar stability following the $M(\rho_c)$ method \cite{Harrison}, where $dM/d\rho_c =0$ represents a turning point from stability to instability. A stability analysis through adiabatic radial oscillations will be left for a future work.

\begin{acknowledgments}
JMZP acknowledges Brazilian funding agency CAPES for PhD scholarship 331080/2019. A. Pradhan thanks to IUCCA, Pune, India for providing facilities under associateship programmes.
\end{acknowledgments}\

%%%%%%%%%%%%%%%%%%%%%%%%%%%%%%%%%%%%%%%%%%%%%%%%%%%%%%%%%%%%%%%%%%%%%%%%%%%%%%%%%%%%

\end{document}